\documentclass[12pt]{article}
\usepackage{graphics,graphicx}
\usepackage{amssymb,epsfig,amsmath,euscript,array}
\usepackage{cite}
\usepackage{pstricks}
\usepackage{color}

\makeatletter \@addtoreset{equation}{section} \makeatother

\newcounter{multieqs}

\newcommand{\be}{\begin{equation}}
\newcommand{\ee}{\end{equation}}

\newcommand{\bm}[1]{\mbox{\boldmath $#1$}}

\newcommand{\kslash}{k \!\!\! / }

\newcommand{\lslash}{l \!\! / }
\newcommand{\Pslash}{P \!\!\!\! / }

\newcommand{\islash}{i \!\!\! / }
\newcommand{\jslash}{j \!\!\! / }
\newcommand{\aslash}{a \!\!\! / }
\newcommand{\bslash}{{b \hspace{-6pt} \slash} }

\newcommand{\onslash}{1 \!\!\! / }
\newcommand{\twslash}{2 \!\!\!/ }
\newcommand{\thslash}{3 \!\!\!/ }
\newcommand{\foslash}{4 \!\!\! / }
\newcommand{\fislash}{5 \!\!\! / }

\newcommand{\mslash}{m \!\!\! / }

\def\bd{\begin{document}}
\def\ed{\end{document}}
\def\nn{\nonumber}
\def\bea{\begin{eqnarray}}
\def\eea{\end{eqnarray}}

\def\red{\color{red}}
\def\black{\color{black}}
\def\blue{\color{blue}}
\def\orange{\color{orange}}

\def\ab{(ijab)}
\def\ba{(ijba)}
\def\ijab{{\tr}_{+}(\islash\, \jslash\, \aslash \, \bslash)}
\def\ijba{{\tr}_{+}(\islash\, \jslash\, \bslash \, \aslash)}
\def\ijaP{{\tr}_{+}(\islash\, \jslash\, \aslash \, \Pslash)}
\def\ijPLa{{\tr}_{+}(\islash\, \jslash\, \Pslash_L \, \aslash)}
\def\ijaPL{{\tr}_{+}(\islash\, \jslash\, \aslash \, \Pslash_L)}
\def\ijPLza{{\tr}_{+}(\islash\, \jslash\, \Pslash_{L;z} \, \aslash)}
\def\ijaPLz{{\tr}_{+}(\islash\, \jslash\, \aslash \, \Pslash_{L;z})}
\def\ijPa{{\tr}_{+}(\islash\, \jslash\, \Pslash \, \aslash)}
\def\iaPb{{\tr}_{+}(\islash\, \aslash\, \Pslash \, \bslash)}
\def\ibPa{{\tr}_{+}(\islash\, \bslash\, \Pslash \, \aslash)}
\def\ijPmu{{\tr}_{+}(\islash\, \jslash\, \Pslash \, \mu)}
\def\ibmuP{{\tr}_{+}(\islash\, \bslash\, \mu \, \Pslash)}
\def\ibmua{{\tr}_{+}(\islash\, \bslash\, \mu \, \aslash)}
\def\iamub{{\tr}_{+}(\islash\, \aslash\, \mu \, \bslash)}
\def\jaPb{{\tr}_{+}(\jslash\, \aslash\, \Pslash \, \bslash)}
\def\ijmuP{{\tr}_{+}(\islash\, \jslash\, \mu \, \Pslash)}
\def\ijmum{{\tr}_{+}(\islash\, \jslash\, \mu \, \mslash)}
\def\ijmmu{{\tr}_{+}(\islash\, \jslash\, \mslash \, \mu)}
\def\ijmP{{\tr}_{+}(\islash\, \jslash\, \mslash \, \Pslash)}
\def\iabP{{\tr}_{+}(\islash\, \aslash\, \bslash \, \Pslash)}
\def\ijbP{{\tr}_{+}(\islash\, \jslash\, \bslash \, \Pslash)}
\def\jbPa{{\tr}_{+}(\jslash\, \bslash\, \Pslash \, \aslash)}
\def\ijPb{{\tr}_{+}(\islash\, \jslash\, \Pslash \, \bslash)}
\def\jbmua{{\tr}_{+}(\jslash\, \bslash\, \mu \, \aslash)}

\def\loablt{ {\tr}_{+}(\lslash_1\, \aslash \, \bslash\, \lslash_2)}

 \def\ijlolt{{\tr}_{+}(\islash\, \jslash\, \lslash_1 \, \lslash_2)}
\def\ijltlo{{\tr}_{+}(\islash\, \jslash\, \lslash_2 \, \lslash_1)}
\def\ibloa{{\tr}_{+}(\islash\, \bslash\, \lslash_1 \, \aslash)}
\def\jaltb{{\tr}_{+}(\jslash\, \aslash\, \lslash_2 \, \bslash)}
\def\ialtb{{\tr}_{+}(\islash\, \aslash\, \lslash_2 \, \bslash)}
\def\bltloa{{\tr}_{+}(\bslash\, \lslash_2\, \lslash_1 \, \aslash)}
\def\jbloa{{\tr}_{+}(\jslash\, \bslash\, \lslash_1 \, \aslash)}
\def\ibPb{{\tr}_{+}(\islash\, \bslash\, \Pslash \, \bslash)}
\def\ijltb{{\tr}_{+}(\islash\, \jslash\, \lslash_2 \, \bslash)}

\def\ijloa{{\tr}_{+}(\islash\, \jslash\,  \lslash_1 \, \aslash)}
\def\ijblt{{\tr}_{+}(\islash\, \jslash\,  \bslash \, \lslash_2)}

\def\jakb{{\tr}_{+}(\jslash\, \aslash\, \kslash \, \bslash)}
\def\iakb{{\tr}_{+}(\islash\, \aslash\, \kslash \, \bslash)}

\def\tofo{{\tr}_{+}(\onslash\, \thslash\, \twslash \, \foslash)}
\def\foto{{\tr}_{+}(\onslash\, \thslash\, \foslash \, \twslash)}
\def\tofi{{\tr}_{+}(\onslash\, \thslash\, \twslash \, \fislash)}
\def\fito{{\tr}_{+}(\onslash\, \thslash\, \fislash \, \twslash)}

\def\lrangle#1#2{\langle #1\,#2\rangle}

\def\Li{{\rm Li}}
\def\eps{\epsilon}
\def\epsuv{{\epsilon_{\rm \mbox{\tiny UV}}}}
\let\bm=\bibitem
\let\la=\label

\def\npb#1#2#3{Nucl. Phys. {\bf{B#1}} #3 (#2)}
\def\plb#1#2#3{Phys. Lett. {\bf{#1B}} #3 (#2)}
\def\prl#1#2#3{Phys. Rev. Lett. {\bf{#1}} #3 (#2)}
\def\prd#1#2#3{Phys. Rev. {D \bf{#1}} #3 (#2)}
\def\cmp#1#2#3{Comm. Math. Phys. {\bf{#1}} #3 (#2)}
\def\cqg#1#2#3{Class. Quantum Grav. {\bf{#1}} #3 (#2)}
\def\nppsa#1#2#3{Nucl. Phys. B (Proc. Suppl.) {\bf{#1A}}#3 (#2)}
\def\ap#1#2#3{Ann. of Phys. {\bf{#1}} #3 (#2)}
\def\ijmp#1#2#3{Int. J. Mod. Phys. {\bf{A#1}} #3 (#2)}
\def\rmp#1#2#3{Rev. Mod. Phys. {\bf{#1}} #3 (#2)}
\def\mpla#1#2#3{Mod. Phys. Lett. {\bf A#1} #3 (#2)}
\def\jhep#1#2#3{J. High Energy Phys. {\bf #1} #3 (#2)}
\def\atmp#1#2#3{Adv. Theor. Math. Phys. {\bf #1} #3 (#2)}
\newcommand{\EQ}[1]{\begin{equation} #1 \end{equation}}
\newcommand{\AL}[1]{\begin{subequations}\begin{align} #1 \end{align}\end{subequations}}
\newcommand{\SP}[1]{\begin{equation}\begin{split} #1 \end{split}\end{equation}}
\newcommand{\ALAT}[2]{\begin{subequations}\begin{alignat}{#1} #2 \end{alignat}
                        \end{subequations}}
\def\beqa{\begin{eqnarray}}
\def\eeqa{\end{eqnarray}}
\def\beq{\begin{equation}}
\def\eeq{\end{equation}}
\def\sst{\scriptscriptstyle}
\def\thetabar{\bar\theta}
\def\Tr{{\rm Tr}}
\def\one{\mbox{1 \kern-.59em {\rm l}}}
 \def\Nh{\hat{N}}

\newcommand{\half}{{\textstyle {1 \over 2}}}

\def\a{\alpha}      \def\da{{\dot\alpha}}
\def\b{\beta}       \def\db{{\dot\beta}}
\def\c{\gamma}  \def\G{\Gamma}  \def\cdt{\dot\gamma}
\def\d{\delta}  \def\D{\Delta}  \def\ddt{\dot\delta}
\def\e{\epsilon}        \def\vare{\varepsilon}
\def\f{\phi}    \def\F{\Phi}    \def\vvf{\f}
\def\h{\eta}
\def\k{\kappa}
\def\l{\lambda} \def\L{\Lambda}
\def\m{\mu} \def\n{\nu}
\def\o{\omega}
\def\p{\pi} \def\P{\Pi}
\def\r{\rho}
\def\s{\sigma}  \def\S{\Sigma}
\def\t{\tau}
\def\th{\theta} \def\Th{\Theta} \def\vth{\vartheta}
\def\X{\Xeta}
\def\z{\zeta}
\def\de{\partial}

\def\cA{{\cal A}} \def\cB{{\cal B}} \def\cC{{\cal C}}
\def\cD{{\cal D}} \def\cE{{\cal E}} \def\cF{{\cal F}}
\def\cG{{\cal G}} \def\cH{{\cal H}} \def\cI{{\cal I}}
\def\cJ{{\cal J}} \def\cK{{\cal K}} \def\cL{{\cal L}}
\def\cM{{\cal M}} \def\cN{{\cal N}} \def\cO{{\cal O}}
\def\cP{{\cal P}} \def\cQ{{\cal Q}} \def\cR{{\cal R}}
\def\cS{{\cal S}} \def\cT{{\cal T}} \def\cU{{\cal U}}
\def\cV{{\cal V}} \def\cW{{\cal W}} \def\cX{{\cal X}}
\def\cY{{\cal Y}} \def\cZ{{\cal Z}}

\def\ua{\underline{\alpha}}
\def\ub{\underline{\phantom{\alpha}}\!\!\!\beta}
\def\uc{\underline{\phantom{\alpha}}\!\!\!\gamma}
\def\um{\underline{\mu}}
\def\ud{\underline\delta}
\def\ue{\underline\epsilon}
\def\una{\underline a}\def\unA{\underline A}
\def\unb{\underline b}\def\unB{\underline B}
\def\unc{\underline c}\def\unC{\underline C}
\def\und{\underline d}\def\unD{\underline D}
\def\une{\underline e}\def\unE{\underline E}
\def\unf{\underline{\phantom{e}}\!\!\!\! f}\def\unF{\underline F}
\def\unm{\underline m}\def\unM{\underline M}
\def\unn{\underline n}\def\unN{\underline N}
\def\unp{\underline{\phantom{a}}\!\!\! p}\def\unP{\underline P}
\def\unq{\underline{\phantom{a}}\!\!\! q}
\def\unQ{\underline{\phantom{A}}\!\!\!\! Q}
\def\unH{\underline{H}}

\def\As {{A \hspace{-6.4pt} \slash}\;}
\def\bs {{b \hspace{-6.4pt} \slash}\;}
\def\Ds {{D \hspace{-6.4pt} \slash}\;}
\def\ds {{\del \hspace{-6.4pt} \slash}\;}
\def\ss {{\s \hspace{-6.4pt} \slash}\;}
\def\ks {{ k \hspace{-6.4pt} \slash}\;}
\def\ps {{p \hspace{-6.4pt} \slash}\;}
\def\pas {{{p_1} \hspace{-6.4pt} \slash}\;}
\def\pbs {{{p_2} \hspace{-6.4pt} \slash}\;}
\def\Ps {{P \hspace{-6.4pt} \slash}\;}
\def\Qs {{Q \hspace{-6.4pt} \slash}\;}

\def\Fh{\hat{F}}
\def\Vh{\hat{V}}
\def\Xh{\hat{X}}
\def\ah{\hat{a}}
\def\xh{\hat{x}}
\def\yh{\hat{y}}
\def\ph{\hat{p}}
\def\xih{\hat{\xi}}
\def\psit{\tilde{\psi}}
\def\Psit{\tilde{\Psi}}
\def\tht{\tilde{\th}}
\def\lt{\tilde{\lambda}}
\def\hl{\hat{\lambda}}
\def\hlt{\hat{\tilde{\lambda}}}
\def\llt{\tilde{l}}
\def\At{\tilde{A}}
\def\Qt{\tilde{Q}}
\def\Rt{\tilde{R}}
\def\Nt{\tilde{N}}

\def\at{\tilde{a}}
\def\st{\tilde{s}}
\def\ft{\tilde{f}}
\def\pt{\tilde{p}}
\def\qt{\tilde{q}}
\def\vt{\tilde{v}}
\def\nt{\tilde{n}}

\def\delb{\bar{\partial}}
\def\bz{\bar{z}}
\def\bD{\bar{D}}
\def\bB{\bar{B}}

\def\bk{{\bf k}}
\def\bl{{\bf l}}
\def\bp{{\bf p}}
\def\bq{{\bf q}}
\def\br{{\bf r}}
\def\bx{{\bf x}}
\def\by{{\bf y}}
\def\bR{{\bf R}}
\def\bV{{\bf V}}

\def\d{\delta}\def\D{\Delta}\def\ddt{\dot\delta}
\def\pa{\partial} \def\del{\partial}
\def\xx{\times}
\def\uno{\mbox{1 \kern-.59em {\rm l}}}
\def\trp{^{\top}}
\def\inv{^{-1}}
\def\dag{{^{\dagger}}}
\def\pr{^{\prime}}
\def\lan{\langle}
\def\ran{\rangle}
\def\rar{\rightarrow}
\def\lar{\leftarrow}
\def\lrar{\leftrightarrow}
\newcommand{\0}{\,\!}      
\def\one{1\!\!1\,\,}
\def\im{\imath}
\def\jm{\jmath}
\newcommand{\tr}{\mbox{tr}}
\newcommand{\slsh}[1]{/ \!\!\!\! #1}
\def\vac{|0\rangle}
\def\lvac{\langle 0|}
\def\hlf{\frac{1}{2}}
\def\ove#1{\frac{1}{#1}}
\def\Box{\square}
\def\ZZ{\mathbb{Z}}
\def\CC#1{({\bf #1})}
\def\bcomment#1{}
\def\bfhat#1{{\bf \hat{#1}}}
\def\VEV#1{\left\langle #1\right\rangle}
\newcommand{\ex}[1]{{\rm e}^{#1}} \def\ii{{\rm i}}
\def\rr{{\rm r}} \def\rs{{\rm s}}\def\rv{{\rm v}}
\def\ri{{\rm i}}\def\rj{{\rm j}}
\newcommand{\lrbrk}[1]{\left(#1\right)}
\newcommand{\sfrac}[2]{{\textstyle\frac{#1}{#2}}}

\font\mybb=msbm10 at 12pt
\def\bb#1{\hbox{\mybb#1}}

\font\myBB=msbm10 at 18pt
\def\BB#1{\hbox{\myBB#1}}

\begin{document}

\begin{flushright}
IPPP/10/XX, DCPT/10/XX
\end{flushright}

\vspace{6pt}

\begin{center}

\hspace{-0.8cm}{\Large \bf Analytic Results for MHV Wilson Loops}

\vspace{15pt}
\end{center}

\centerline{\mbox {\large Paul~Heslop$^{a}$ and
Valentin~V.~Khoze$^{b}$}%
{
\renewcommand{\thefootnote}{}  \footnotetext{
{\tt \{paul.heslop, valya.khoze\}@durham.ac.uk} } } }

\begin{center}
{\small \em
\begin{itemize}
\item[\ \ \ \ \ \ $^a$]
Institute for Particle Physics Phenomenology,  \\
Department of Mathematical Sciences and Department of Physics\\
Durham University,
Durham, DH1 3LE, United Kingdom\\
\item[\ \ \ \ \ \ $^b$]
Institute for Particle Physics Phenomenology,  \\
Department of Physics,
Durham University, \\
Durham, DH1 3LE, United Kingdom

\end{itemize}
}


\vspace{23pt} {\bf Abstract}
\end{center}

\noindent We obtain concise analytic formulae for Wilson loops
computed on special $n$-point polygonal contours
through two-loops in weakly coupled ${\mathcal N}=4$ supersymmetric gauge theory. The contours we consider
can be embedded into a $(1+1)$-dimensional subspace of the
4-dimensional gauge theory, corresponding to the boundary of the
$AdS_3$ on the string theory side. Our analytic results hold for any number
of edges, thus generalising to arbitrary $n$ the recently derived expressions for 2-dimensional octagons.
These polygonal Wilson loops have been conjectured to be
equivalent to MHV scattering amplitudes in planar ${\mathcal N}=4$ SYM.

\setcounter{page}{0} \thispagestyle{empty}
\newpage


\section{Introduction}
\label{sec:1}
\setcounter{footnote}{0}

An ambitious goal of solving the ${\mathcal N}=4$ supersymmetric Yang-Mills theory (SYM)
should include a full understanding of the structure of dynamical quantities in this theory, in particular
the S-matrix, or scattering amplitudes.

It was shown in \cite{am} that at strong coupling, scattering amplitudes
can be determined through a calculation of a Wilson loop
\begin{equation}
\label{wil}
W_n \, :=\, W[ \cC_n]  \ = \ {\rm Tr} \, \cP \exp \left[ i g\oint_{\cC_n} \! d\tau  \ \dot{x}^{\mu} (\tau )A_\mu (x(\tau ))   \right]
\ ,
\end{equation}
with a light-like $n$-edged polygonal contour $\cC_n$ obtained by  attaching the momenta of the scattered particles
$p_1, \ldots , p_n$ one after the other, following the order
of the colour generators in the colour-ordered scattering amplitude. The vertices, $x_i$, of the polygon are related to the
external momenta via $p_i= x_i-x_{i+1},$ where $x_{n+1}=x_1$.

Since then it has been conjectured that at any value of the
coupling in planar ${\mathcal N}=4$ SYM there is a non-trivial
relation between scattering amplitudes and Wilson loops~
\cite{am,dks,bht}. The weak coupling relation between MHV
scattering amplitudes and Wilson loops has been supported by an
increasing amount of
evidence~\cite{am,dks,bht,dhks4,dhks5,seven,dhks6,Anastasiou:2009kn}.
In perturbation theory the number of different Feynman topologies
in the Wilson loop does not grow with the number of external
particles (edges) $n$ \cite{Anastasiou:2009kn} and is much smaller
than the corresponding number of integrals for the $n$-point
amplitude~\cite{vergu}. This, together with the fact that the
integrals themselves are also more straightforward to compute,
makes the computation of $W_n$ considerably more attractive
practically than that of the amplitudes.

In practice, the duality between MHV amplitudes and Wilson loops beyond one-loop is understood in terms of the remainder function.
The remainder function of the amplitude
${\cR}_n$ (or of the Wilson loop ${\cR}^{WL}_n$) is defined as the
difference between the logarithm of the amplitude $\cM_n$
(or Wilson loop $W_n$) and the known BDS expression obtained
in~\cite{bds,abdk}, so that
\be
\label{eq:1}
  {\cR}_n\,=\,\log (\cM_n) -(BDS)_n \ , \qquad
  {\cR}^{WL}_n\,=\,\log (W_n) -(BDS)^{WL}_n\, .
\ee
Here $\cM_n$ is the colour-ordered MHV amplitude normalised by the tree-level result,
$\cM_n := \cA_n^{\rm MHV}/\cA_n^{\rm MHV\, tree}$
The Wilson loop possesses an anomalous conformal symmetry and the
remainder function constitutes the correctly regularised, conformally
invariant part of the Wilson loop, reducing the
number of independent variables down to the conformally invariant
cross-ratios~\cite{dhks5}.
The Wilson loop/amplitude duality states that
the two remainder functions are
identical~\cite{seven,dhks6,Anastasiou:2009kn},
${\cR}_n=  {\cR}^{WL}_n.$

Ref. \cite{Anastasiou:2009kn} has assembled a general algorithm for computing Wilson loops $W_n$ for arbitrary $n$
at two loops and has studied their multi-collinear limits.
The actual computations in \cite{Anastasiou:2009kn} were carried out numerically for up to $n=8$ edges in general kinematics, and in
\cite{bhkt,hk} for up to $n=30$ edges for special families of regular polygons.
Fully analytic two-loop calculations of $W_n$ were pioneered for the case of the hexagon in \cite{6an1,6an2}
and derived using quasi-multi-Regge kinematics at intermediate stages to simplify the computation.
The result at $n=6$ was expressed in terms of a 17-pages long linear combination of generalised polylogarithm
functions of uniform transcendental weight four (or alternatively it was also recast in \cite{Zhang:2010tr} in terms of one-dimensional integrals.)
Following these exciting developments, the authors of~\cite{Goncharov:2010jf} were able to find another representation
of the hexagon result at two-loops in terms of a remarkably simple compact expression involving only $\Li_m$ functions with
$m\le 4$ and logarithms (again of total weight four).

Another recent inspiring achievement stems from the study of Wilson loops whose contours are chosen to lie
in a $(1+1)$-dimensional subspace of Minkowski space-time. At strong coupling, this corresponds to Wilson loops
embedded into the boundary of $AdS_3$ target space of the dual string theory. Alday and Maldacena in \cite{am8}
have computed the corresponding Wilson loop at strong coupling using an auxiliary integrable system, and
have expressed their result in terms of a simple one-dimensional integral.
In Ref.~\cite{bhkt} octagonal Wilson loops in the same kinematics were evaluated numerically also at weak coupling
and it was suggested that there are pronounced similarities between the weak and the strong coupling contributions in this
special 2-dimensional kinematics. The analytic result for $(1+1)$-dimensional octagon at two loops was very recently
derived by the authors of \cite{dds8} following the technology established in their earlier papers \cite{6an1,6an2}.
The striking result of \cite{dds8} is that all the complications related to multiple occurrences of generalised polylogarithms
have mutually cancelled in the final expression for the $(1+1)$-dimensional octagon, leaving only a product of four logarithms
plus constant terms in the two loop expression for the Wilson loop remainder function.
The striking simplicity of this analytic end result for the $(1+1)$-dimensional octagon makes it even simpler than the
corresponding strong coupling result derived in \cite{am8} (even though numerically the two functions remain very close).

The motivation of this paper is to investigate whether wider classes
of Wilson loops exist to the eight point case
of~\cite{dds8}, but which exhibit a similarly simple analytic form
at weak coupling. We will show that this is
indeed the case. We shall consider Wilson loops with an arbitrary number
$n$ of light-like edges and will require that they can be embedded into a $(1+1)$-dimensional subspace of Minkowski space.
These Wilson loops are also conformally equivalent to those with contours in $(2+1)$-dimensions whose spatial
projection circumscribes the unit circle \cite{am8}.
For these Wilson loops we will find that all the polylogarithms
appearing at one- and two-loops disappear from final answers and  the
entire two loop contribution is described by a weight-four function composed entirely of logarithms (and constant terms).

Our programme is in two parts. First we construct an analytic expression for
$\log W_n$ which is required to satisfy certain precise criteria. Then
we verify that this analytic expression does agree with the numerical evaluation
of $\log W_n$ which we carry out following the algorithm of \cite{Anastasiou:2009kn}.
Since our numerical computations can be carried out at a press of a button for any
kinematics and any $n$-polygons and with an ever increasing accuracy, this is where
the programme of numerically evaluating Wilson loops (or remainders) starts paying off.

To come up with the analytic ansatz in the first place, we will employ the following strategy.
Firstly we assume that only logarithms can appear and none of the complicated polylogarithms are
allowed in $\log W_n$ for the $(1+1)$-dimensional contours.
Secondly we assume that the arguments of these logarithms for the conformally-invariant part of the
answer\footnote{This means the $\Li_2$ part of the one-loop answer and the remainder function
$\cR_n$ at two loops.}
are the cross-ratios themselves, and the total expression at each loop-level
has the appropriate uniform transcendental weight.

We then start assembling various pieces of evidence for
the resulting ansatz, based on simplifying the known one-loop expressions at all $n$ of \cite{bddk}
together with employing the
recently found eight-point two-loop result of~\cite{dds8}.
Further evidence suggesting
the appearance of simple cross-ratios only as arguments of the
logarithms (and not more complicated
functions of cross-ratios) can be found  from the recent tremendously
simplified form of the six-point remainder function \cite{Goncharov:2010jf}. At first sight
this result seems to suggest the opposite, since  there one finds
highly complicated functions (involving square roots) of the simple cross-ratios appearing as
arguments of the polylogarithms. However these complicated functions
were interpreted in terms of momentum twistors. In the special
kinematics in $(1+1)$-dimensions the momentum twistors go to simple
products of space-time coordinates, and so this observation
also points to the fact that only cross-ratios themselves should
appear as arguments in the specialised kinematics.

This paper is organised as follows. In section \ref{sec:2} we introduce the two-dimensional kinematics
 and define the cross-ratios relevant to polygonal contours with $n$ light-like edges.
 In section \ref{sec:3} we show that Wilson loops in this kinematics
 depend only on simple logarithms at one loop by simplifying the one-loop
 expression. In section \ref{sec:4}
 we move on to the discussion of two-loop results
 at $n=8$, $n=10$ and $n=12$ points. The corresponding analytic expressions are given in
 Eqs.~\eqref{r8},\eqref{r10} and \eqref{r12-old}.
 We then construct the general analytic expression valid for any $n$ in \eqref{rn}.
 All of these results satisfy stringent tests from the multi-collinear limits and are verified numerically
 following the methods developed in \cite{Anastasiou:2009kn}.
 Finally we also check consistency of these results with the regular polygons computations performed earlier in \cite{bhkt}.
 We present our conclusions in section \ref{sec:5} where we also comment on the possible structure of higher loop contributions
 beyond two loops.
An Appendix discusses the consistency of our two loop expression with multi-collinear limits in more detail.

\section{Two-dimensional kinematics}
\label{sec:2}

The remainder function is a conformally-invariant object. As such it depends on the kinematics only through
conformally-invariant cross-ratios \cite{dks,dhks5}.
In this section we will define these cross-ratios in the $(1+1)$-dimensional case. This is
relevant for polygonal contours with $n$ light-like edges which can be embedded into the boundary
of the three-dimensional anti-de-Sitter subspace, $AdS_3$ of the dual
string theory  $AdS_5 \times S^5$ target space.

\begin{figure}[h!]
\begin{center}
\includegraphics[width=7cm]{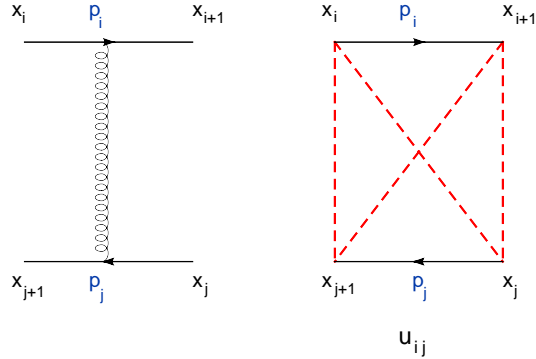}
\end{center}
\caption{\it The left figure shows the one-loop Wilson loop
diagram which gives the finite part of the two-mass easy box
function with massless momenta $p_i,p_j$ as in \cite{bht}.  On the
right we represent the
  corresponding cross-ratio
  $u_{ij}$,  the red dashed lines depicting the factors $x_{ij}^2$,  $x_{i+1 j+1}^2$, $x_{i+1j}^2$, $x_{ij+1}^2$ in
  the definition of $u_{ij}$ in~\eqref{uijdef}. Later we will
  represent the cross-ratio as  a single line stretched between edges
  $i$ and $j$ similarly to the gluon propagator.}
\label{fig:cross-ratio}
\end{figure}

For a polygonal contour with $n$ light-like edges, in general,
there are $n(n-5)/2$ independent conformal cross-ratios (if we do not,
as in \cite{Anastasiou:2009kn}, impose the Gram determinant
constraints). This is the same as the number as of two-mass easy boxes. As always, we use the basis for the
cross ratios $u_{i,j}$,
\begin{equation}\label{uijdef}
  u_{ij}={x_{ij+1}^2 x^2_{i+1 j} \over x_{ij}^2 x_{i+1 j+1}^2} \ ,
\end{equation}
which `connect' edges $i$ and $j$, as shown in
Figure~\ref{fig:cross-ratio}. Here $x_i$ are the vertices of the
polygonal contour of the Wilson loop, and $j \ge i+3$ modulo $n$.

For the Wilson loop contour to be embeddable into two space-time
dimensions\footnote{The Wilson loop itself of course is calculated
in $D=4-2\epsilon$ dimensions, only its contour is now embedded
into $(1+1)$-dimensions which can be thought of as the boundary of
$AdS_3$.} the number of edges $n$ must be even.
In two dimensions the number of independent cross-ratios reduces and they
have to satisfy the following conditions,
\begin{eqnarray}
u_{i\,,  i+ {\rm odd}} &=& 1 \nonumber \\
u_{2i+1\,,   2j+1} &=& u^+_{ij} \qquad    2 \leq (i-j)\, \, {\rm mod} \, n/2  \le n/2-2    \label{eq:AdS3} \\
u_{2i\,,  2j} &=& u^-_{ij}\nonumber
\end{eqnarray}
where we have defined
the two dimensional cross-ratios $u^\pm_{ij}$ in terms of
light-cone coordinates as will be explained momentarily. Clearly, just
as the indices of the cross-ratios $u_{ij}$ are defined mod $n$, the
two-dimensional light-cone cross-ratios $u_{ij}^\pm$ have indices
defined mod $n/2$.
The
sequence of $(1+1)$-dimensional light-like edges of the contour
must have a zig-zag form as in \cite{am8}, so that any edge pointing forward in time
must be followed by an edge pointing backwards in time, otherwise the consecutive
$(1+1)$-dimensional edges would be indistinguishable from one
another. This implies that the vertices of the contour have the
following simple light-cone representation:
\begin{align}
  x_{2i}=(x_i^+,x_i^-)\ , \qquad   x_{2i+1}=(x_i^+,x_{i+1}^-)\ , \qquad i=1,\ldots, n \,.
\end{align}

\begin{figure}[h!]
    \centering
      \includegraphics[width=9cm]{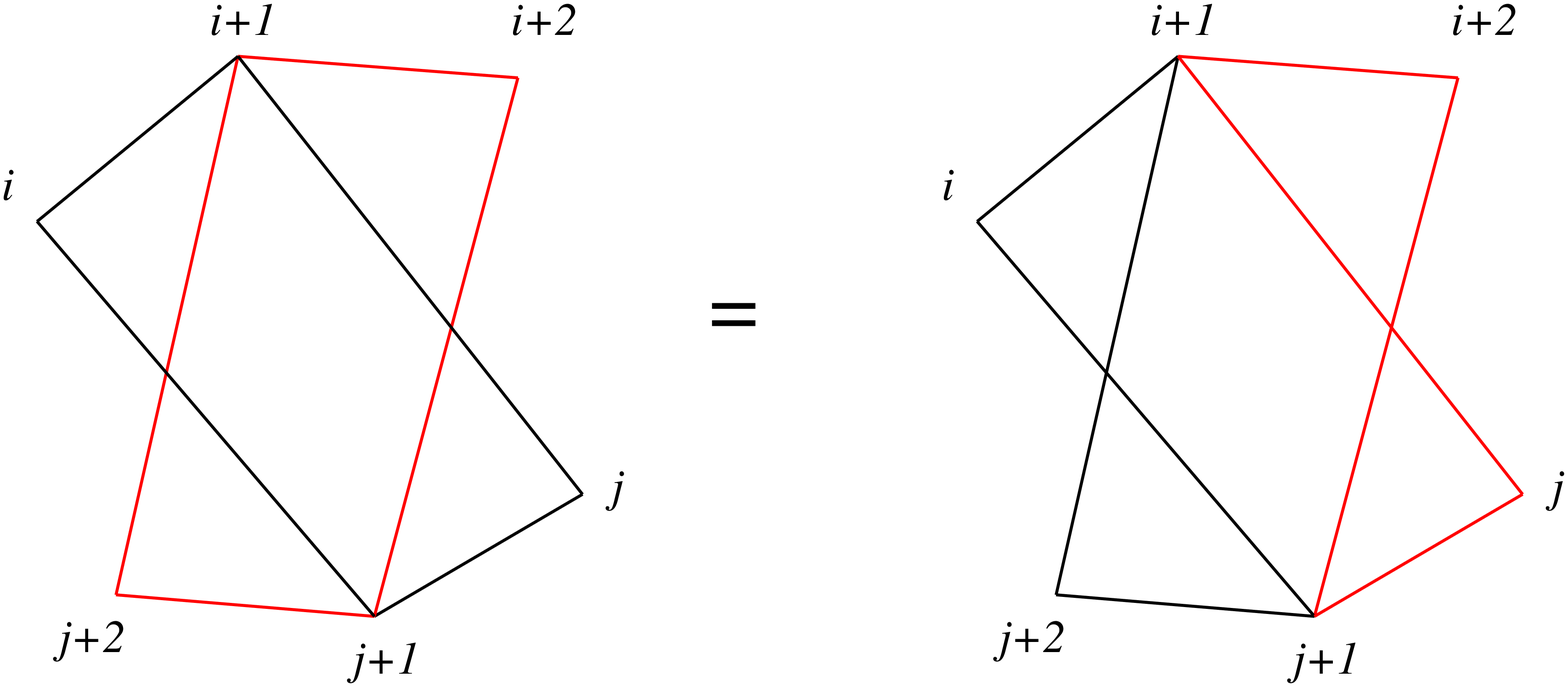}
    \caption{Figure illustrating equation~(\ref{ueq}). On the
      left we represent in black the rectangular
      cross-ratio $(1- 1/u^\pm_{i\, j})$ and in red the rectangular cross-ratio $(1-
1/u^\pm_{i+1\, j+1}).$ On the right, on the other hand, in black
we show  the ``crossed" cross-ratio $1-u_{i j+1}$ and in red
$1-u^\pm_{i+1 j}$. Clearly the product on the left-hand side
equals that on the right-hand side.}
    \label{fig:upm}
  \end{figure}

The cross-ratios $u_{ij}^{\pm}$ appearing on the right hand side
of \eqref{eq:AdS3} are functions of only either $x^+$ or $x^-$
light-cone coordinates, they are defined via,
\begin{equation}\label{uijdef2}
  u_{ij}^{+}\, :=\,{x^+_{ij+1}\, x^+_{i+1 j} \over x^+_{ij}\, x^+_{i+1 j+1}} \ , \qquad
   u_{ij}^{-}\, :=\,{x^-_{ij+1}\, x^-_{i+1 j} \over x^-_{ij}\, x^-_{i+1 j+1}} \ ,
\end{equation}
and as such, these cross-ratios are essentially made from
one-dimensional distances.  It is easy to check that this results
in the following simple identity
\begin{align}
\label{ueq}
  (1-u^\pm_{i\, j+1})   (1-u^\pm_{i+1\, j})& \,=\,   (1- 1/{u^\pm_{i\,
      j}})  (1-1/{u^\pm_{i+1\, j+1}})\\
       u_{i, i+1}=u_{i+1, i}&=0 \qquad u_{i,i }=\infty  \ .
\end{align}
This identity can be verified directly. It can also be understood
geometrically as follows from Figure~\ref{fig:upm}.

Also note
that this equation is precisely  the $AdS_3$ Y-system equation
of~\cite{agm}, where the $Y$'s of~\cite{agm} (evaluated at
$\zeta=0$) are associated with the cross-ratios as
\begin{align}
  u_{k,-k-1}^+ = {Y_{2k} \over 1+Y_{2k}} \qquad \qquad u_{k,-k-2}^- =
  {Y_{2k+1} \over 1+Y_{2k+1}}\ .
\end{align}
We will thus refer to~(\ref{ueq}) as the Y-system from now on.

\section{Wilson loops for two-dimensional kinematics depend on simple logs only }
\label{sec:3}

Here we start assembling evidence that there is an
enormous simplification for Wilson loops in $D=4-2\epsilon$
dimensions when their polygonal contours are restricted to two
space-time dimensions. The conformal cross-ratios in the
two-dimensional kinematics were defined in the previous sections.
We will see below that Wilson loops in this kinematics will
contain only logarithmic functions (at least at one and two
loops). In particular, all the dependence on polylogarithms and
other more complicated functions disappears.

The first piece of evidence occurs at one loop.
The fact that the BDS expression can be simplified to depend only on logarithms
in $AdS_3$ kinematics has previously been observed by Alday and Maldacena
in~\cite{am8}. We shall re-derive this result, exhibiting on the way a few useful
identities which we have discovered.

The general all $n$ expression for the one-loop result for the Wilson loop/ MHV amplitude \cite{bddk,bht} is well-known.
It is given as follows
for $n>4$,
\begin{align}\label{m1}
\cM_n^{(1)}(\epsilon)&= -{1\over 2 \e^2}\sum_{i=1}^n   \left( - {t_i^{[2]}
    \over \mu_\mathrm{}^2 }\right)^{-\e} +\ F_n^{(1)}(\epsilon)\ ,\\
F_n^{(1)}(0)& = {1 \over 2} \sum_{i=1}^n g_{n,i} \ ,
\end{align}
where
\begin{equation}
g_{n,i}  \ = \
-\sum_{r=2}^{[ n/2 ]  -1}
  \ln \left( { -t^{[r]}_{i}\over -t^{[r+1]}_{i} }\right)
  \ln \left({ -t^{[r]}_{i+1}\over -t^{[r+1]}_{i} }\right) \, + \,
D_{n,i} \, + \, L_{n,i} + {3\over 2} \zeta_2 \  ,
\end{equation}
and  $t^{[r]}_{i} := (p_i + \cdots + p_{i+r-1})^2$
are the kinematical invariants.  The functions $D_{n,i}$ and $L_{n,i}$
for the case at hand with even number of edges,
$n=2m$, are given by,
\begin{eqnarray}
\label{LDeven}
D_{2m,i} &=& -\sum_{r=2}^{m-2}
\Li \left( 1- { t^{[r]}_{i} t^{[r+2]}_{i-1}
\over t^{[r+1]}_{i} t^{[r+1]}_{i-1} }  \right)
- {1 \over 2} \Li \left( 1- { t^{[m-1]}_{i} t^{[m+1]}_{i-1}
\over t^{[m]}_{i} t^{[m]}_{i-1}} \right) \ ,
\\ \nonumber
L_{2m,i} &=& - {1\over 4}
  \ln \left({ -t^{[m]}_{i}\over -t^{[m]}_{i+m+1}  } \right)
  \ln \left({ -t^{[m]}_{i+1}\over -t^{[m]}_{i+m} } \right) \ .
\end{eqnarray}

As is well-known, the finite part of the one-loop result is
of transcendental weight two and it contains both $\log^2$ terms and dilogarithms. The total contribution of
all dilogs in the one-loop expression is,
\begin{align}
  \sum_{\{u\}} \Li_2  \left( 1- u \right)
\end{align}
where the sum is over all cross-ratios $u_{ij}$.

At eight points in  two-dimensional kinematics the cross-ratios
always appear in pairs $u$ and  $1-u$. Indeed at
eight-points the Y-system equation~(\ref{ueq}) becomes simply
\begin{align}
{\rm Octagon:} \qquad \quad  (1-1/u^\pm_{ij})(1-1/u^\pm_{i+1\,j+1})=1
\end{align}
giving $u^\pm_{i+1\,j+1}=1-u^\pm_{ij}$.
This can also be seen from the explicit form
for the octagon cross-ratios
in the standard parametrisation \cite{am8,bhkt} in terms of $\chi^+$ and $\chi^-$ variables,
\begin{eqnarray}
\label{eightcrosses}
{\rm Octagon:} \qquad \quad
 u_{15} &=& {\chi^{+}\over 1 + \chi^{+}} :=u^+_{24}\ , \qquad u_{26} \ = \ {\chi^{-}\over 1 + \chi^{-}}:=u^-_{13}
\ ,
\\ \nonumber
 u_{37} &=& {1\over 1 + \chi^{+}}:=u^+_{13} \ , \qquad u_{48} \ = \ {1\over 1 + \chi^{-}}:=u^-_{24}\ ,
\\ \nonumber\\ \nonumber
u_{i\,, i+3} &=& 1\ ,  \ \  \qquad \qquad i=1, \ldots , 8
\ ,
\end{eqnarray}
which immediately gives $u_{15}=1-u_{37}$ and $u_{48}=1-u_{26}.$

Then a dilog identity of Euler
\begin{align}
  \text{Li}_2(u)=-\text{Li}_2(1-u)-\log (1-u) \log (u)+\frac{\pi ^2}{6}
\end{align}
allows us to rewrite the one loop Wilson loop purely in terms of
logarithms~\cite{am8}.

Remarkably,  similar cancellations of dilogs occur at higher points too,
albeit with the use of much more
complicated identities.

We have found  numerically, with the help of the PSLQ
algorithm~\cite{PSLQ}, that the following general identity holds for
any even $n$:
\begin{align}\label{genpolyid}
\sum _{i=1}^{n/2} \sum _{j=i+2}^{{n/2}-\delta_{i,1}} \text{Li}_2(1-u^\pm_{i,j})+
  \sum _{i=1}^{n/2} \sum _{j=i+1}^{n/2} \sum _{k=j+1}^{n/2} \sum _{l=k+1}^{n/2}
    \log u^\pm_{i,k} \log u^\pm_{j,l} - \pi^2 {({n/2}-6) \over 12}\,=\, 0
\end{align}
for any $u_{ij}^\pm>0$, with indices defined mod $n/2$ and satisfying the Y-system equation~(\ref{ueq}).

This identity then kills all logarithms in the one loop
$n$-point Wilson loop/MHV amplitude over a two-dimensional
contour. (Note that the first term in
~(\ref{genpolyid}) is simply an explicit writing of the sum over all
(plus or minus) cross-ratios
and is equivalent to~(\ref{LDeven})).

To illustrate we take the next simplest example, $n=10$. Here there are 25 cross-ratios.
\begin{eqnarray}
\label{tencrosses}
{\rm Decagon:} \qquad \quad
&&u_{i\,, i+3} \,=\, 1\ ,   \qquad i=1, \ldots , 10\ ,
\\ \nonumber
 &&u_{i,i+4}  \ , \qquad\qquad i=1,\ldots,10
\ ,
\\ \nonumber
&&u_{i\,, i+5} \,=\, 1\ ,   \qquad i=1, \ldots , 5
\ .
\end{eqnarray}
The $u_{i\,, i+4}$ group of 10 can be divided into two groups of 5
cross-ratios $u^\pm_{i\, i+2}$ as in~(\ref{eq:AdS3}) which we here
define simply as $u^\pm_i$ to simplify the notation slightly.
Each of these groups separately satisfies the constraint,
\begin{align}\label{con}
  1 - u^\pm_i = (1 - 1/u^\pm_{i+2}) (1 - 1/u^\pm_{i+3}) \qquad i=1\dots 5\ ,
\end{align}
which
arises directly from the Y-system~(\ref{ueq}).

In this case the identity~(\ref{genpolyid}) becomes
\begin{align}
\sum_{i=1}^5 \Big(\text{Li}_2\left(1-u_i\right) +\log
   \left(u_i\right) \log \left(u_{i+1}\right)\Big)-\frac{\pi ^2}{3}=0
\end{align}
for any $u_i$ which satisfy \eqref{con}.
The indices in the above two equations are all understood to be Mod 5.

Note that the fact that the one-loop contribution to the Wilson loop depends only on logarithms
in $AdS_3$ kinematics has previously been observed
in~\cite{am8} where they find that the BDS expression in these
special kinematics simply reduces to (e.g. see Eq.~(E.2) of Ref.~\cite{am8})
\begin{align}
\label{AMBDSlog}
 - { 1 \over 4 } \sum_{i=1}^{n/2}\,  \sum_{ j=1,   j \not = i,i-1 }^n  \log { x^+_{ji} \over x^+_{j+1,i} }
 \log { x^-_{j ,i-1} \over x^-_{j ,i }}\ .
\end{align}

\section{Two-loop results}
\label{sec:4}

 At two loops the remainder function for $(1+1)$-dimensional octagons has recently
 been shown to be independent of polylogarithms taking a strikingly simple form~\cite{dds8}
\begin{align}\label{r8DDS}
  R_{8}^{\rm DDS}= -\frac{1}{2} \log(1+\chi^+) \log(1+\frac{1}{\chi^+}) \log(1+\chi^-) \log(1+\frac{1}{\chi^-})
\, -\, \frac{\pi^4}{18}\ .
\end{align}
To prepare for the higher-$n$ analysis we recast this expression in terms of logarithms of cross-ratios,
$u_i:=u_{i i+4}$,
\begin{align}\label{r8}
  R_{8}^{\rm DDS}= -\frac{1}{2} \log(u_1) \log(u_2) \log(u_3) \log(u_4)
\, -\, \frac{\pi^4}{18}\ .
\end{align}

 Knowing (from the previous section and from \cite{am8}) that at one-loop dilogs do cancel for all $n$-points,
 and that there are no polylogarithms at two-loops at $n=8$ points
 it is natural to ask whether this simplicity
 extends to higher points at two loops.

At $n=8$ points the form of the octagon remainder \eqref{r8} is essentially fixed
by multi-collinear limits and (cyclic/parity) symmetry up to a single factor, once one
makes an educated guess that the final result only depends on logs of cross-ratios.

We found that conditions imposed by multi-collinear limits together with this logs-only assumption are even
stronger for higher points, completely fixing the form of higher-point Wilson loops.
We will now demonstrate this explicitly for
the 10-point and 12-point Wilson loops .

\subsection{Decagon}

Define $u_i:=u_{i i+4}$.

Our result for the decagon Wilson loop remainder function at 2-loops and in the $(1+1)$-dimensional
kinematics is
\begin{align}\label{r10}
  R_{10}= -\frac{1}{2} \Big(\log(u_1) \log(u_2) \log(u_3) \log(u_4) \,+\, {\rm cyclic}\Big)
\, -\, \frac{\pi^4}{12}\ ,
\end{align}
where `cyclic' means that the first term with labels $1,2,3,4$ is accompanied by
$2,3,4,5$ plus eight more terms up to $10,1,2,3.$

The expression above is fixed by the triple collinear limit which gives
very strong constraints.\footnote{Triple
collinear limits in terms of remainder functions were discussed in detail in \cite{Anastasiou:2009kn}.
In general, as explained in \cite{hk}, in the limit where $k+1$ consecutive momenta (edges) become collinear, the remainder function
transforms as $\cR_{n} \to \cR_{n-k} + \cR_{k+4}$ where the second term on the {\it r.h.s.} arises from the
corresponding splitting function.}
For concreteness take  momenta $p_8$, $p_9$ and $p_{10}$
to be collinear.  Under this triple collinear limit $\cR_{10} \to \cR_{8} + \cR_{6}$ and we
have that $u_5\rightarrow 1,\ u_7 \rightarrow 0, \ u_9 \rightarrow
1$. Moreover, the cross-ratios $u_1,u_2,u_3$ of $\cR_{10}$ coincide with those of $\cR_8$,
and the combination $u_4 u_{10}$ maps onto $u_4$ of $\cR_8$.

In this limit the remainder function should reduce as follows
\begin{align}
  R_{10}(u_i) \rightarrow& R_8( u_1,u_2,u_3,u_4 u_{10}) +R_6 \\
= &-\frac{1}{2} \Big(\log(u_1) \log(u_2) \log(u_3) \log(u_4 u_{10}) \Big)
-\frac{\pi^4}{12}\ .
\end{align}
This is quite a strong constraint. In particular there should be no
dependence on the variables $u_6, u_8$ and the cross-ratios  $u_4$ and
$u_{10}$ must appear only in the combination $u_4 u_{10}$.

The above putative remainder~(\ref{r10}) has  precisely these properties.

We have also verified numerically, using the technology developed in \cite{Anastasiou:2009kn},
that our analytic expression \eqref{r10} is in precise agreement with the numerical results.

\subsection{Dodecagon}

Here there are 42 cross-ratios,
\begin{eqnarray}
\label{twelvcrosses}
{\rm Dodecagon:} \qquad \quad
&&u_{i\,, i+3} \,=\, 1\ ,   \qquad i=1, \ldots , 12\ ,
\\ \nonumber
 &&u_{i,i+4} \,:=\, u_i\,, \qquad i=1,\ldots,12
\ ,
\\ \nonumber
&&u_{i\,, i+5} \,=\, 1\, ,   \qquad i=1, \ldots , 12
\ ,
\\ \nonumber
&&u_{i\,, i+6} \,:=v_i\, ,   \qquad i=1, \ldots , 6
\ .
\end{eqnarray}

In the triple collinear limit with collinear $p_{10}$, $p_{11}$ and $p_{12},$ we have $u_7\rightarrow 1,\ u_9
\rightarrow 0,\ u_{11}  \rightarrow 1, v_5 \rightarrow 1$ and the remainder function
should reduce as
\begin{align}
  R_{12}(u_i;v_i) \rightarrow R_{10}(u_1,u_2,u_3,u_4,u_5,u_6 v_6, v_1,
  v_2,v_3,v_4 u_{12} ) + R_6\ .
\end{align}
As previously, all dependence on $ u_8$ and   $ u_{10}$ is lost and the
dependence on $u_6,v_6,v_4 , u_{12} $ appears only via $u_6 v_6$ and $v_4 u_{12}$.

The 12 point Wilson loop which satisfies these properties is
\begin{align}
 R_{12}= -1/2 \Big(&
  \log(u_1) \log(u_2) \log(u_3) \log(u_4) + 11 \ \mathrm{cyclic}\nonumber\\
&+  \log(u_1) \log(u_2) \log(u_3) \log(v_4) + 11 \ \mathrm{cyclic}\nonumber\\
&+ \log(u_1) \log(u_2) \log(v_3) \log(v_4) + 11 \ \mathrm{cyclic}\nonumber\\
&+ \log(u_1) \log(v_2) \log(v_3) \log(v_4) + 11 \ \mathrm{cyclic}\nonumber\\
&+ \log(v_1) \log(v_2) \log(v_3) \log(v_4) + 5 \ \mathrm{cyclic}\Big)\,-\, \pi^4/9\ .
\label{r12-old}
\end{align}

There is another collinear limit one can consider on $R_{12}$, the
quintuple collinear limit. Under this limit we have
\begin{align}
  u_5 \rightarrow 1, \quad u_{11}\rightarrow 1, \quad v_3 \rightarrow 1,
  \quad v_5 \rightarrow 1, \quad v_1 \rightarrow 0 \ ,
\end{align}
and the remainder should split as
\begin{align}\label{r12quint}
  R_{12}(u;v) \rightarrow R_8(u_1,u_2,u_3,u_4 v_4 v_{12})+ R_8(u_7,
  u_8,u_9,u_{10}v_4 u_{6}) \ .
\end{align}
This is a highly non-trivial check of our function $R_{12}$ and we
stress that no information
from this quintuple collinear limit was used in the determination of
$R_{12}$. Nevertheless, as one can easily check, the
function $R_{12}$ defined in (\ref{r12-old}) satisfies~(\ref{r12quint}) in
this collinear limit.

Finally, we have checked that
the analytic expression \eqref{r12-old} agrees well with a sample of numerical data points which we computed for $n=12$.

\subsection{General $n$-point at two loops}

\begin{figure}[h]
    \centering
\includegraphics[width=8cm]{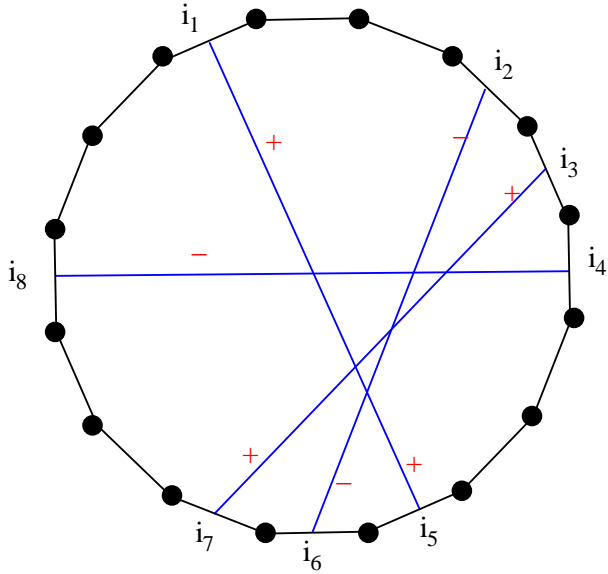}
    \caption{The two loop $n$-point remainder function is given in~\eqref{rn}.
    It consists of a sum of terms of the form
    $\log ( u_{i_1 i_5}) \log ( u_{i_2 i_6}) \log ( u_{i_3 i_7}) \log ( u_{i_4 i_8})$
    one term of which is represented pictorially here. One sums over all possible ways of
    drawing four mutually crossing lines connecting edges of the polygon, such that the parity of the lines alternates.}
\label{fig:rn}
  \end{figure}
The above results at 8, 10 and 12 points are special cases of the
following formula for general $n$
\begin{align}\label{rn}
  R_n &= -{1 \over 2} \Big( \sum_{\cS} \log ( u_{i_1
    i_5}) \log ( u_{i_2 i_6}) \log ( u_{i_3 i_7}) \log ( u_{i_4 i_8})
  \Big) - {\pi^4\over 72} (n-4)\ ,
\end{align}
where the sum runs over the set
\begin{align}
\cS &= \Big\{ i_1, \dots i_8: 1\leq i_1<i_2< \dots < i_8 \leq n,\qquad  i_k
-i_{k-1} = \mathrm{odd} \Big\}\ .
\end{align}
This sum has the following geometrical interpretation. Represent
the cross-ratios $u_{ij}$ as lines from edge $i$ to $j$ of our
polygonal contour and assign a label $+$ or $-$ to the line
corresponding to whether $i$, $j$ are both odd or both even
(corresponding to whether it is a $+$ or $-$ cross-ratio in $(1+1)$-dimensions
-- see~(\ref{eq:AdS3}).)
The sum is then over all
ways of drawing four mutually crossing lines connecting edges of the
contour, with the parity of the lines alternating (see figure~\ref{fig:rn} ).

Under the triple collinear limit in which edges
$n-2,n-1, n$ become collinear (and in which in fact $n-1$ becomes soft)
one has
\begin{align}
  u_{i, n-1}\rightarrow 1, \qquad u_{1,n-3} \rightarrow 0 \ ,
\end{align}
and the remainder function should reduce as
\begin{align}\label{gencol}
  R_n(u_{ij})\rightarrow R_{n-2}(\hat u_{ij} ) + R_{6}
\end{align}
where the $n-2$-point cross-ratios $\hat u_{ij}$ are defined in terms
of the $n$-point cross-ratios in the collinear limit as
\begin{align}
  \hat u_{i \, n-2} = u_{i \, n-2}  u_{i \, n} \qquad \hat u_{ij} =
  u_{ij} \ i,j \neq n-2 \ .
\end{align}

Indeed one can check that the above formula~\eqref{rn} does indeed
reduce precisely as required by~\eqref{gencol}.

We should also check the correct behaviour under more general collinear
limits. In the case where the $2k+1$ edges $n-2k, \dots,  n$ become
collinear we have
\begin{align}\label{collimit}
u_{1,n-2k-1}\rightarrow 0, \quad   u_{i, n-2k+1} \rightarrow 1, \quad   \dots \quad   u_{i, n-1} \rightarrow 1  \quad \quad (2 \leq
  i \leq n-2k-2)\ ,
\end{align}
and the remainder function should reduce as
\begin{align}
  R_n(u_{ij}) \rightarrow R_{n-2k}(\hat u_{ij} )   +R_{2k+4}(u'_{ij})
\end{align}
where the cross-ratios of the reduced remainders  are related to the
$n$-point cross-ratios as
\begin{align}\label{collimit2}
\hat u_{i,n-2k} = u_{i, n-2k} \dots u_{i,n},  \quad \hat u_{ij}=u_{ij}
\qquad
1\leq i,j < n-2k \\
u'_{i,2} = u_{i, 2} \dots u_{i,n-2k-2},  \quad u'_{ij}=u_{ij} \qquad
0\geq i,j \geq -2k \ .
\end{align}
We have checked explicitly (using a computer) in numerous non-trivial examples
that our remainder function~\eqref{rn} satisfies all the correct
multi-collinear limits.
In the appendix we explain with the help of some pictures why the
multi-collinear limit works.

\subsection{Comparison with the $n$-point regular polygons}

Using the general formula for the $n$-point remainder we can now
specialise to the $Z_n$-symmetric or regular polygons studied previously at strong
coupling~\cite{am8} and numerically at weak coupling~\cite{bhkt}.
The cross-ratios for the regular polygons take the form
as in~\cite{bhkt}
\begin{eqnarray}
\label{regpolcrra}
u_{ij} & = & 1 \ ,   \qquad \qquad \qquad \qquad \ i\, - \,j  \, = \, {\rm odd} \ ,
\nonumber \\
u_{ij} & = & 1\, - \, \left(
{
\sin {2 \pi\over   n}
\over
\sin {\pi (i-j) \over   n}
} \right)^2
\ , \qquad
i\, - \,j  \, = \, {\rm even} \ .
\end{eqnarray}
Since these regular polygons can be embedded into $(1+1)$-dimensions (up to a conformal transformation),
the cross-ratios in \eqref{regpolcrra} are of the form required by \eqref{eq:AdS3}
and explicitly solve
the Y-system equation~(\ref{ueq}).

Plugging these into the general remainder
function~(\ref{rn}) gives us the $n$-point remainder function for the
regular polygons. Table~\ref{table2} displays these results against
the corresponding numerical results computed
in~\cite{bhkt}. One can see that they perfectly agree
to 3 digits.
\begin{table}[ht]
\begin{center}
\scalebox{0.75}[0.75]{
\begin{tabular}{|c||c|c|c|c|c|c|c|c|c|c|}
\hline $n$ & $8$ & $10$ & $12$ & $14$ & $16$ & $ 18 $ & $20$& $22$ & $24$ & $30 $
\\
\hline \hline $\mathcal{R}^\mathrm{num}_{n} $   &  -5.528  &  -8.386 & -11.262 & -14.145 & -17.035 &-19.926 & -22.821& -25.717& -28.614& -37.311
\\
 $\mathcal{R}^\mathrm{an}_{n} $&
-5.52703 & -8.38554 & -11.2606 & -14.1444 & -17.0334 & -19.9257 &
   -22.8204 & -25.7167 & -28.6142  & -37.3114\\
\hline
\end{tabular}
}
\end{center}
\caption{\it Numerical values for the regular $n$-point Wilson loop 2-loop remainder function against the
  corresponding analytic values obtained from our general analytic
  formula~(\ref{rn}).}
\label{table2}
\end{table}
This agreement provides additional evidence in favour of our main result~(\ref{rn}).

The regular polygons considered above live in $(2+1)$-dimensions
and are obtained from special $(1+1)$-dimensional polygons with a conformal transformation. They are regular
in the sense that they possess the discrete $Z_n$ symmetry in 2 spatial dimensions.
One can also consider other less restricted examples of polygons with discrete symmetries, such as $Z_{n/4}$-symmetric polygons (for $n$ divisible by 4). All such cases provide convenient settings
for numerical verification of the general analytic expression~(\ref{rn}).

\section{Conclusions}
\label{sec:5}

Based on a combination of analytic and numerical techniques as well as on the use of multi-collinear limits,
we derived a compact analytic formula for Wilson loops with contours embedded in $(1+1)$-dimensional subspace
of Minkowski space.
Our two-loop expression for the remainder function \eqref{rn}
holds for any number
of external momenta (or edges), thus generalising to arbitrary $n$ the recent result for 2-dimensional octagons computed in \cite{dds8}.

In our view, it is a striking feature of the all-$n$ results presented here
that while the direct analytic NLO computation
(when available) {\it at the first instance} gives a very complicated answer, the end result \eqref{rn} is
a compact logs-only expression.
One expects that there must be an alternative simpler way to compute weakly coupled Wilson loops
or indeed the scattering amplitudes themselves. It is known that at strong coupling there is an integrable theory set-up
at work \cite{am8,agm,Hatsuda:2010cc,amsv,Alday:2010ku}. It would be interesting to find an appropriate alternative
formalism applicable at weak coupling as well.

At the same time, the simplicity of one-loop and two-loop results
strongly suggests that similarly simple formulae (involving only
logarithms) should apply to three-loops and beyond in special
kinematics. In particular a simple generalisation of~(\ref{rn})
involving six logarithms is immediately apparent. Diagrammatically one
simply writes down six mutually crossing lines, alternately even and
odd (ie joining even and odd sides) in all possible ways. This
automatically satisfies all possible  collinear limits (when  an
appropriate $n$ dependent constant is added) as can be seen from
similar arguments to those in the appendix
and has been checked extensively using a computer.
Indeed an  $L$-loop generalisation is now  apparent also, simply
involving $4L$ logarithms.
However, such a three loop expression will vanish at 8 and 10 points
(and the $L$ loop generalisation will only be non-zero from $4L$ points)
as there are not enough edges to write it down. If these were the only
possible $L$-loop structures, then this would mean
that the all orders octagon remainder in $1+1$ dimensions would be equal to the
two loop expression (up to multiplicative and additive, coupling
dependent constants.) We now know that this is not the case (and so there
must be
additional structures occurring at higher loops on top of these)
since the analytic
expression at strong coupling~\cite{am8} does not have the
same form as at weak coupling~\cite{dds8}. Intriguingly,  however
the strong coupling result does agree  numerically to quite high
accuracy with the two loop one~\cite{bhkt}.

\vspace{.3cm}

\section*{Acknowledgements}

It is a pleasure to thank Claude Duhr for many enlightening
discussions and insights at early stages of this work. We also thank  Andreas
Brandhuber, Patrick Dorey, Mark Spradlin and Gabriele
Travaglini for discussions.
VVK acknowledges a Leverhulme Research Fellowship.

\newpage




\section*{Appendix: multi-collinear limits}
\label{sec:multi-coll-limits}

In this appendix we explain in some detail why our $n$-point 2 loop
expression correctly reproduces the multi-collinear limits.
 We illustrate this in the case of the
quintuple collinear limit acting on the 12 point Wilson loop, but the
discussion generalises in a straightforward way to an arbitrary
multi-collinear limit acting on an arbitrary polygon in a
straightforward manner. We take
edges $8,9,10,11,12$ (which we collectively refer to as `$A$' to approach the collinear limit, with edges 9
and 11 also becoming soft. In this limit the expected collinear
behaviour is $R_{12} \rightarrow R_8
+R_8$~\cite{Anastasiou:2009kn,hk}
\begin{center}
  \includegraphics[height=3.5cm]{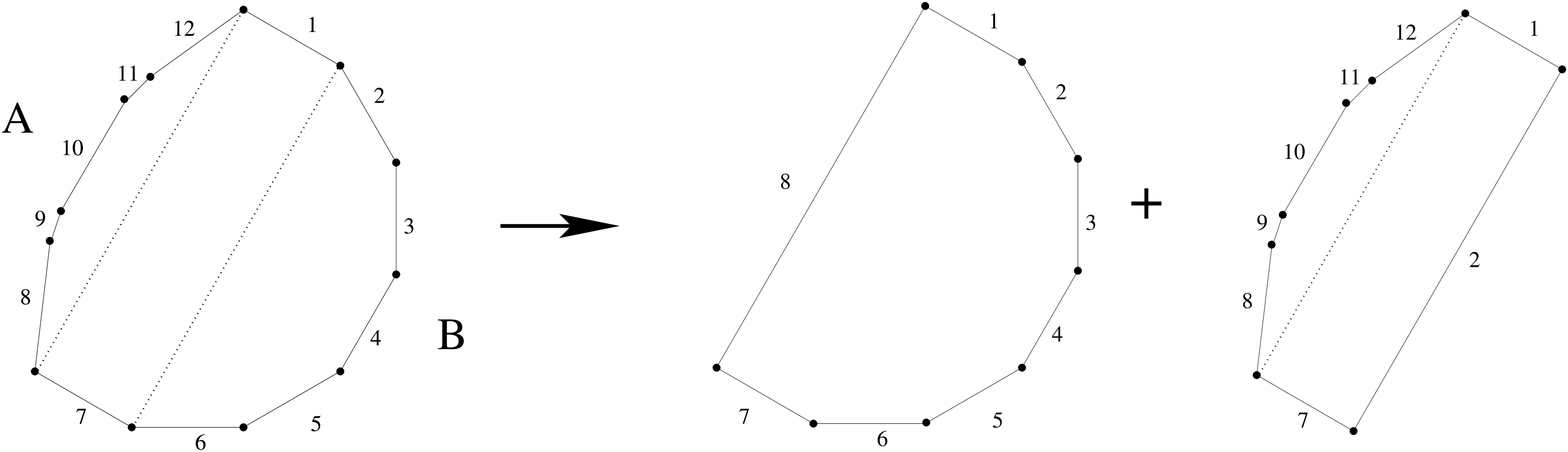}
\end{center}
In this diagram the edges $8,9,10,11,12$ are becoming collinear, approaching the dashed
edge. Note that there is a complete symmetry between this case and that of
taking the edges  $2,3,4,5,6$ (which we will call $B$) collinear. Indeed as shown recently
in~\cite{Alday:2010ku}, a conformal transformation can take us from
one  collinear limit to the other.

How does the $n$-point result~(\ref{rn}) respect this collinear limit?
Let us consider different terms in the expression. As in
figure~\ref{fig:rn} we will represent $\log(u_{ij}$)  by a line drawn
between edge
$i$ and edge $j$. We will call such a line `odd' if
$i,j$ are odd and `even' if they are even
(corresponding to $u_{ij}$ being a + or a - cross-ratio in (1+1)
dimensions, see~(\ref{eq:AdS3})). The two loop Wilson loop expression~(\ref{rn})
then corresponds to a sum over terms with four lines.

From~(\ref{collimit}) we see that any term containin a single odd line between A and B
vanishes in the limit, for example
\begin{center}
  \includegraphics[height=3.5cm]{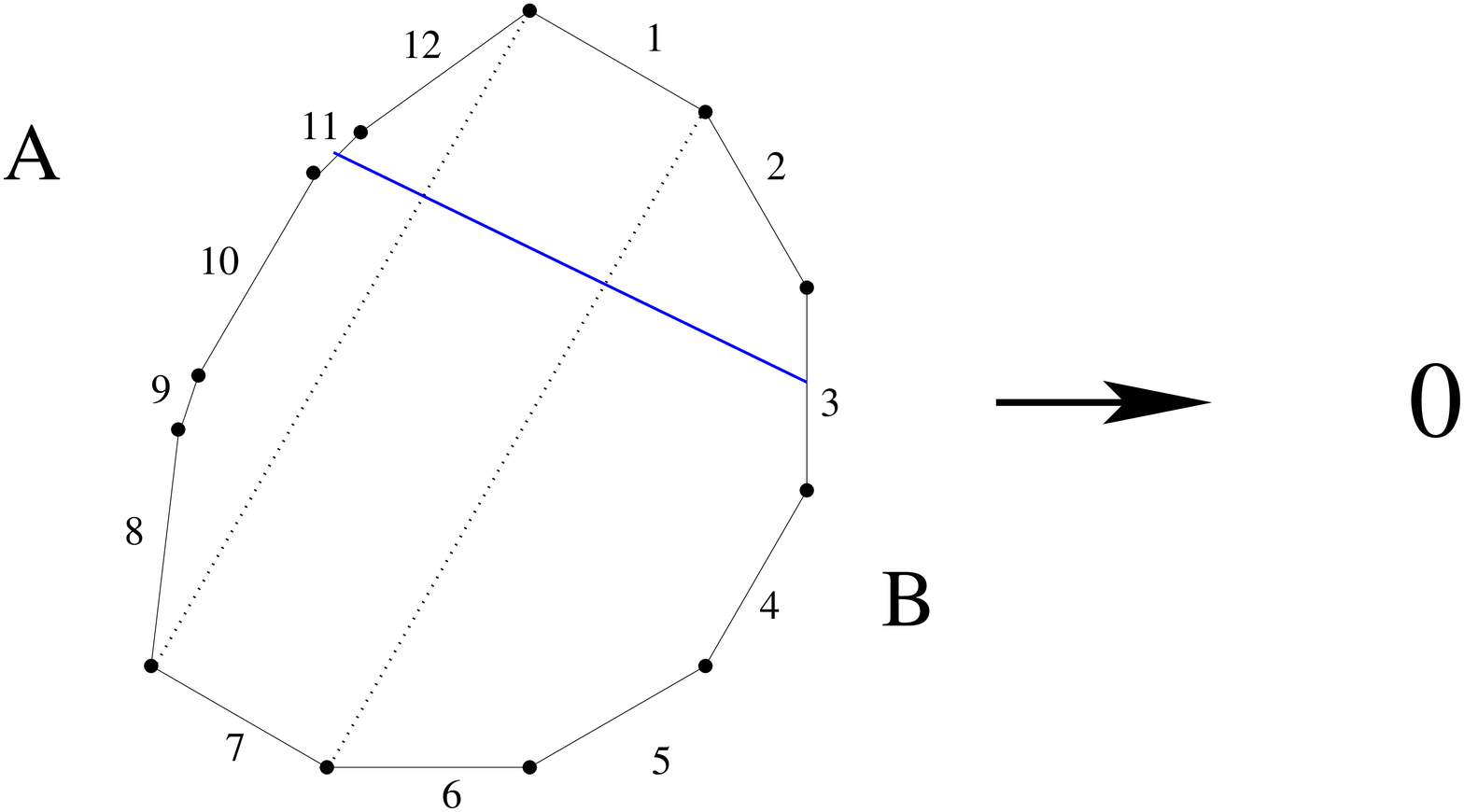}
\end{center}

What can we say about `even' lines between A and B?  Crucially, in any given $\log^4$ term in the $n$-point
expression~(\ref{rn}) we will never have more than one such `even' edge
(between A and B) surviving in the collinear limit. This is because
any given term which contains a product of two `even' lines
between edges $A$
and $B$ must inevitably also  contain  an odd line stretching between edges A and
B, sandwiched between the two `even' lines. The odd line will then
vanish in the collinear limit and hence kill such a term in the
expression.
So `even' lines between A and B occur alone in the collinear limit: the accompanying three
lines must stretch either from $A^c$ to $A^c$ (the complement of $A$)
or stretch from $B^c$ to $B^c$. Indeed the three accompanying lines
must either
{\em all} be in $A^c$ or all in $B^c$ since if one  line was  in $A^c$ and
another in $B^c$ then the lines would not intersect each other.

On the other hand `even' lines between A and B sum up to give the
respective line of the reduced Wilson loop (recall that the relevant reduced
cross-ratio is a product of the
original cross-ratios~(\ref{collimit2}).) So any four lines
involving an `even' line between  A and B must come accompanied by other
similar terms to reproduce four lines in the reduced expression. For
example,
 \begin{center}
  \includegraphics[height=3.5cm]{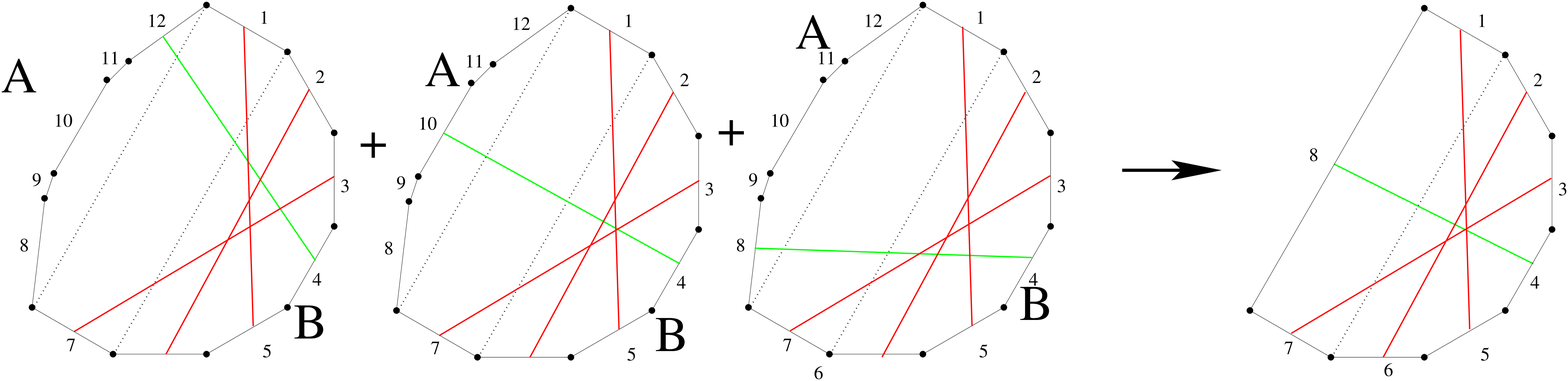}
\end{center}

We have thus covered all situations involving lines between A and B. All
other terms must consist of four lines entirely in $A^c$ or entirely
in $B^c$ (as before we can not have a mixture of such lines as they
would not intersect.) These are not present in our example but can
occur at higher points. These terms simply reproduce corresponding terms in
the reduced Wilson loops. Finally there is  only one more case to
consider, the
single line
between the two edges in $A^c \cap B^c$ (ie between edges $1$ and $7$
in our example.) This term should not occur in the collinear limit,
and indeed this is the case. The reason is that such a line will
inevitably have to be accompanied by three lines between A and B (since
they have to cross each other.) Two of these line will be `even', but
the other one will be odd and hence vanish in the collinear limit.


\begin{thebibliography}{99}


\bibitem{am}
  L.~F.~Alday and J.~Maldacena,
  {\it Gluon scattering amplitudes at strong coupling,}
  JHEP {\bf 0706} (2007) 064,
  {\tt 0705.0303 [hep-th]}.

\bibitem{dks}
  J.~M.~Drummond, G.~P.~Korchemsky and E.~Sokatchev,
  {\it Conformal properties of four-gluon planar amplitudes and Wilson loops,}
  Nucl.\ Phys.\  B {\bf 795} (2008) 385,
  {\tt 0707.0243 [hep-th]}.


\bibitem{bht}
  A.~Brandhuber, P.~Heslop and G.~Travaglini,
  {\it MHV Amplitudes in N=4 Super Yang-Mills and Wilson Loops,}
  Nucl.\ Phys.\  B {\bf 794} (2008) 231,
  {\tt 0707.1153 [hep-th]}.

\bibitem{dhks4}
  J.~M.~Drummond, J.~Henn, G.~P.~Korchemsky and E.~Sokatchev,
  {\it On planar gluon amplitudes/Wilson loops duality,}
  Nucl.\ Phys.\  B {\bf 795} (2008) 52,
  {\tt 0709.2368 [hep-th]}.

\bibitem{dhks5}
  J.~M.~Drummond, J.~Henn, G.~P.~Korchemsky and E.~Sokatchev,
{\it Conformal Ward identities for Wilson loops and a test of the duality with
  gluon amplitudes,}
Nucl.\ Phys.\  B {\bf 826} (2010) 337
  [arXiv:0712.1223 [hep-th]].

\bibitem{seven}
  Z.~Bern, L.~J.~Dixon, D.~A.~Kosower, R.~Roiban, M.~Spradlin, C.~Vergu and A.~Volovich,
{\it The Two-Loop Six-Gluon MHV Amplitude in Maximally Supersymmetric Yang-Mills
  Theory,} Phys.\ Rev.\  D {\bf 78}, 045007 (2008),
  {\tt 0803.1465 [hep-th]}.

\bibitem{dhks6}
  J.~M.~Drummond, J.~Henn, G.~P.~Korchemsky and E.~Sokatchev,
  {\it Hexagon Wilson loop = six-gluon MHV amplitude,}
  Nucl.\ Phys.\  B {\bf 815} (2009) 142,
{\tt 0803.1466 [hep-th].}

\bibitem{Anastasiou:2009kn}
  C.~Anastasiou, A.~Brandhuber, P.~Heslop, V.~V.~Khoze, B.~Spence and G.~Travaglini,
 {\it Two-Loop Polygon Wilson Loops in N=4 SYM,}  JHEP {\bf 0905} (2009) 115,
  {\tt 0902.2245 [hep-th]}.

\bibitem{vergu}
  C.~Vergu,
  {\it The two-loop MHV amplitudes in N=4 supersymmetric Yang-Mills theory,}
  arXiv:0908.2394 [hep-th].



\bibitem{bds}
  Z.~Bern, L.~J.~Dixon and V.~A.~Smirnov,
  {\it Iteration of planar amplitudes in maximally supersymmetric Yang-Mills
  theory at three loops and beyond,}
  Phys.\ Rev.\  D {\bf 72} (2005) 085001,
  {\tt hep-th/0505205}.

\bibitem{abdk}
  C.~Anastasiou, Z.~Bern, L.~J.~Dixon and D.~A.~Kosower,
  {\it Planar amplitudes in maximally supersymmetric Yang-Mills theory,}
  Phys.\ Rev.\ Lett.\  {\bf 91} (2003) 251602,
  {\tt hep-th/0309040}.

\bibitem{bhkt}
  A.~Brandhuber, P.~Heslop, V.~V.~Khoze and G.~Travaglini,
  {\it Simplicity of Polygon Wilson Loops in N=4 SYM,}
  JHEP {\bf 1001}, 050 (2010)
  [arXiv:0910.4898].

\bibitem{hk}
  P.~Heslop and V.~V.~Khoze,
  {\it Regular Wilson loops and MHV amplitudes at weak and strong coupling,}
  JHEP {\bf 1006} (2010) 037
  [arXiv:1003.4405 [hep-th]].

\bibitem{6an1}
  V.~Del Duca, C.~Duhr and V.~A.~Smirnov,
  {\it An Analytic Result for the Two-Loop Hexagon Wilson Loop in N = 4 SYM,}
  arXiv:0911.5332.

\bibitem{6an2}
  V.~Del Duca, C.~Duhr and V.~A.~Smirnov,
  {\it The Two-Loop Hexagon Wilson Loop in N = 4 SYM,}
  arXiv:1003.1702.

\bibitem{Zhang:2010tr}
  J.~H.~Zhang,
  {\it On the two-loop hexagon Wilson loop remainder function in N=4 SYM,}
  arXiv:1004.1606 [hep-th].

\bibitem{Goncharov:2010jf}
  A.~B.~Goncharov, M.~Spradlin, C.~Vergu and A.~Volovich,
  {\it Classical Polylogarithms for Amplitudes and Wilson Loops,}
  arXiv:1006.5703 [hep-th].

\bibitem{am8}
  L.~F.~Alday and J.~Maldacena,
  {\it Null polygonal Wilson loops and minimal surfaces in Anti-de-Sitter space,}
  JHEP {\bf 0911}, 082 (2009)
  [arXiv:0904.0663 [hep-th]].

\bibitem{dds8}
  V.~Del Duca, C.~Duhr and V.~A.~Smirnov,
  {\it A Two-Loop Octagon Wilson Loop in N = 4 SYM,}
  arXiv:1006.4127 [hep-th].


\bibitem{bddk}
Z.~Bern, L.~J.~Dixon, D.~C.~Dunbar and D.~A.~Kosower, {\it One
Loop N Point Gauge Theory Amplitudes, Unitarity And Collinear
Limits,} Nucl.\ Phys.\ B {\bf 425} (1994) 217, {\tt
hep-ph/9403226}.

\bibitem{agm}
  L.~F.~Alday, D.~Gaiotto and J.~Maldacena,
  {\it Thermodynamic Bubble Ansatz,}
  arXiv:0911.4708.


\bibitem{PSLQ}
H. R. P. Ferguson,  and D. H. Bailey,  {\it A Polynomial Time,
  Numerically Stable Integer Relation Algorithm.} RNR
Techn. Rept. {\bf RNR-91-032} (1992).



\bibitem{Hatsuda:2010cc}
  Y.~Hatsuda, K.~Ito, K.~Sakai and Y.~Satoh,
{\it Thermodynamic Bethe Ansatz Equations for Minimal Surfaces in $AdS_3$,}
  arXiv:1002.2941 [hep-th].

\bibitem{amsv}
  L.~F.~Alday, J.~Maldacena, A.~Sever and P.~Vieira,
  {\it Y-system for Scattering Amplitudes,}
  arXiv:1002.2459.

\bibitem{Alday:2010ku}
  L.~F.~Alday, D.~Gaiotto, J.~Maldacena, A.~Sever and P.~Vieira,
  {\it An Operator Product Expansion for Polygonal null Wilson Loops,}
  arXiv:1006.2788 [hep-th].




\end{thebibliography}
\end{document}